\documentclass[twocolumn,prl]{revtex4}

\usepackage{graphicx}
\usepackage{dcolumn}
\usepackage{amsmath}

\begin{document}
\preprint{HEP/123-qed}

\title[Short Title]{
Coherent
Lattice Vibrations in
Carbon Nanotubes}

\author{Y.~S.~Lim,$^{1,2}$ K.~J.~Yee,$^{3}$ J.~H.~Kim,$^{3}$
J.~Shaver,$^{1,5}$ E.~H.~Haroz,$^{1,5}$
J.~Kono,$^{1,5,\dagger}$\\ S.~K.~Doorn,$^{4}$
R.~H.~Hauge,$^{5}$} \author{R.~E.~Smalley$^{5}$}

\affiliation{$^{1}$Department of Electrical and Computer
Engineering, Rice University, Houston, Texas
77005, USA\\ $^{2}$Department of Applied Physics, Konkuk
University, Chungju, Chungbuk 380-701, Republic
of Korea\\ $^{3}$Department of Physics, Chungnam
National University, Daejeon 305-764, Republic of
Korea\\ $^{4}$Division of Chemistry, Analytical
Chemistry Science, Los Alamos National
Laboratory, Los Alamos, New Mexico 87545, USA\\
$^{5}$Carbon Nanotechnology Laboratory, Rice
University, Houston, Texas 77005, USA}

\date{\today}

\begin{abstract}
We have generated and detected the radial-breathing
mode of coherent lattice vibrations in single-walled carbon
nanotubes using ultrashort laser pulses. Because the band gap is
a function of diameter, these diameter oscillations cause
ultrafast band gap oscillations, modulating interband excitonic
resonances at the phonon frequencies (3-9~THz).
Excitation spectra show a large number of pronounced peaks,
mapping out chirality distributions in great detail.
\end{abstract}

\pacs{aaa}
\maketitle

Electrons, phonons, and their mutual interaction determine most
of the properties of crystalline solids~\cite{Ziman60Book}.
Optical and electrical properties, in particular, are almost
entirely dominated by these two fundamental excitations and it
is the subtle interplay between them that gives rise to phenomena
such as the Franck-Condon principle and superconductivity.  With
the advent of ultrafast spectroscopy, one can probe electronic
and vibrational dynamics in real
time~\cite{Shah99Book,Merlin97SSC}.   Single-walled carbon
nanotubes (SWNTs), with their uniquely-simple crystal structures
and chirality-dependent electronic and vibrational states,
provide a one-dimensional playground for studying the dynamics
and interactions of electrons and phonons.  Recent CW optical
studies of
SWNTs~\cite{OconnelletAl02Science,BachiloetAl02Science,ZaricetAl04Science,DoornetAl04APA,TelgetAl04PRL,FantinietAl04PRL,SfeiretAl04Science,WangetAl05Science,MaultzschetAl05PRB,JorioetAl05PRB,MaultzschetAl05PRB2}
have produced a world of intriguing phenomena, including
phonon-assisted photoluminescence, strongly bound excitons, and
chirality-dependent resonant Raman scattering (RRS) -- results
of interaction between excited electronic/excitonic states and
phonons~\cite{PerebeinosetAl05PRL,ThomsenReich06Chapter,PopovetAl04NL,MachonetAl05PRB,Goupalov1,JiangetAl05PRB,Goupalov2}.

Here we show
a real-time observation of lattice vibrations in SWNTs. Using
pump-probe spectroscopy we observed coherent phonons (CPs) in
individual SWNTs, corresponding to the radial breathing mode
(RBM).  The observed RBMs were found to correspond exactly to
those seen by CW RRS for the same sample but with narrower phonon
linewidths, no photoluminescence signal or Rayleigh scattering
background to obscure features, and excellent resolution
allowing normally blended peaks to appear as distinct features.
Additionally, differences in the RBM intensity of families of
SWNTs were observed between CP and RRS.  Finally, when viewing
the excitation profiles of several RBMs, we found a two-peak
resonance, which we attribute to this technique acting as a
modulation spectroscopy; i.e., we are observing the first
derivative of the interband excitonic absorption peak.

The sample used in this study was a micelle-suspended SWNT
solution with a diameter range of 0.7-1.3~nm.  The SWNTs were
individually suspended with sodium dodecyl sulfate in
D$_2$O~\cite{OconnelletAl02Science}  in a quartz cell with an
optical path length of 1~mm.
We performed degenerate pump-probe measurements at room
temperature using $\sim$50~fs pulses at a repetition rate of
89~MHz from a
mode-locked Ti:Sapphire laser
with an average pump power $\sim$20~mW.
We tuned the center wavelength in 5-nm steps from 710~nm to
860~nm (1.75-1.43~eV) by controlling the slit between the
intracavity prism pair in the Ti:Sapphire laser.
In the CW RRS experiments on the same sample, the excitation
source was a
Ti:Sapphire laser with a power of 15~mW at the sample.
Signal collection was done using a
triple monochromator and
CCD camera.

\begin{figure}
\includegraphics [scale=0.85] {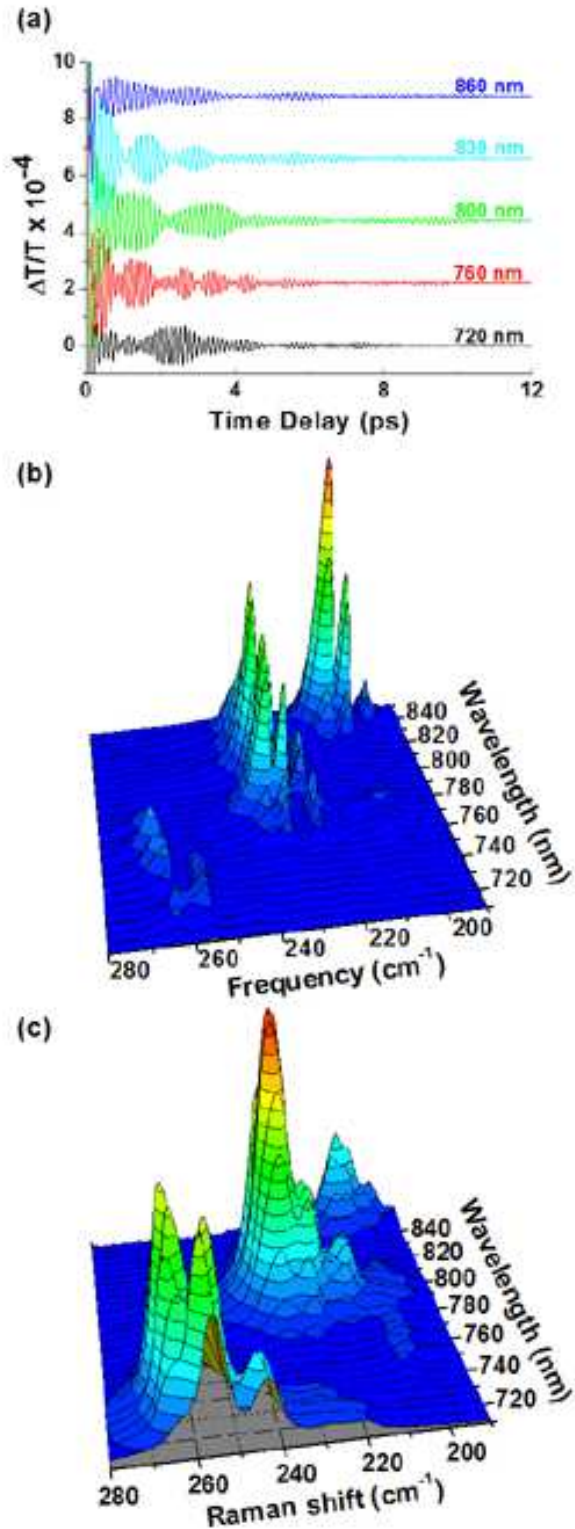}
\caption{(a) Coherent phonon oscillations excited and measured
at five different photon energies.
(b) A 3D plot of the Fast Fourier Transform of coherent phonon
oscillations obtained over a photon energy range of 710-850~nm
(1.746-1.459~eV) with a 5-nm step size.  (c) A 3D plot of
resonance Raman scattering over an excitation energy range of
710-850~nm (1.746-1.459~eV) with a 5-nm step size.} \end{figure}

Figure 1(a) shows coherent phonon oscillations in SWNTs excited
at different pump photon energies.  The amplitude of
oscillations in terms of normalized differential transmission
was $\sim$10$^{-4}$ near zero delay.  Each
trace consists of a superposition of multiple oscillation modes
with different frequencies, exhibiting a strong beating pattern.
 The beating pattern sensitively changes with the photon
energy, implying that the CP oscillations are dominated by RBMs
which are resonantly enhanced by pulses commensurate with their
unique electronic transitions, as is the case in CW Raman
scattering.  The dephasing time of the dominant CP oscillations
was $\sim$5~ps.

To determine the frequencies of the excited lattice vibrations,
we took a fast Fourier transform (FFT) of the time-domain
oscillations to produce CP spectra [Fig.~1(b)].  For comparison,
we show CW RRS spectra [Fig.~1(c)] for the same sample.  The CP
spectra [Fig.~1(b)] are seen to cluster into three distinct
regions, similar to CW RRS~\cite{DoornetAl04APA}, as shown in
Fig.~1(c).  The main peak positions coincide between Fig.~1(b)
and Fig.~1(c), indicating that the oscillations seen in Fig.~1(a)
are indeed due to the RBM of coherent lattice vibrations.
However, upon close examination, we find a few noticeable
differences between the CP data and CW RRS data: i)~unresolved
shoulder features in RRS are seen as resolved peaks in the CP
spectra due to narrower linewidths, ii)~there are different intensity distributions among the
strongest peaks in the three distinct regions, and iii)~the CP
spectra show a surprising double-peak dependence on the photon
energy.  These differences are discussed in more depth in
the following.

\begin{figure}
\includegraphics [scale=0.93] {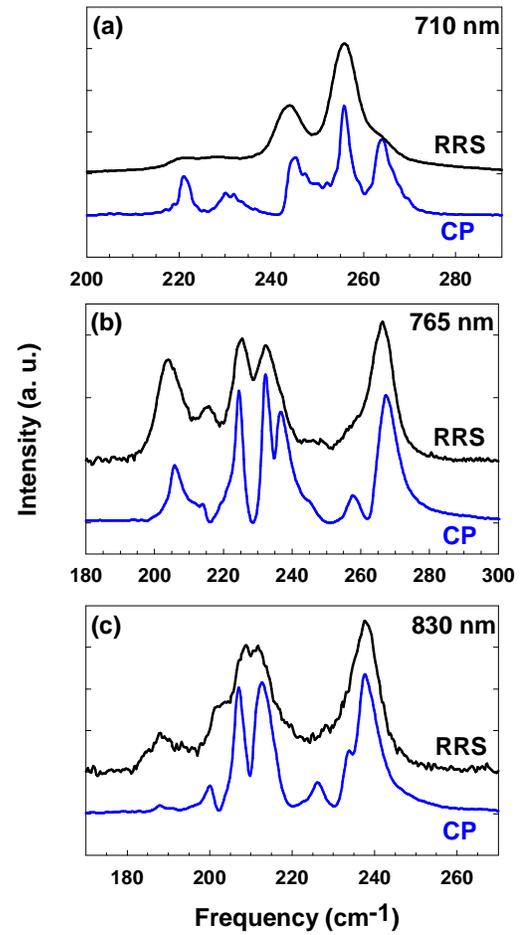}
\caption{Phonon spectra using three different photon
energies obtained from resonant Raman scattering (RRS) and
coherent phonon (CP) measurement.  Coherent phonon
data exhibits dramatically narrower linewidths.
} \end{figure}

Figures 2(a)-2(c) present direct comparison between CP spectra
and CW RRS spectra.  Here, the CP spectra obtained with
different wavelengths are overlaid on the equivalent RRS spectra
taken at the same wavelengths of 710~nm (1.746~eV), 765~nm
(1.621~eV), and 830~nm (1.494~eV), respectively.  There is
overall agreement between the two spectra in each figure.
However, it is clear that many more features are resolved
in the CP data.  The narrower linewidths in the CP data make it
possible to resolve blended peaks in the RRS data.  With less
overlap among the close-by peaks, more precise determination of
line positions is possible.  Through peak fitting using
Lorentzians, we have identified and successfully assigned 18
RBMs in the CP spectra over the 1.44-1.75~eV (710-860~nm) photon
energy range.

The measured average linewidth of RBMs of semiconducting
nanotubes obtained from CP measurements is
approximately 3~cm$^{-1}$.   Those measured in this study for
HiPco material using CW RRS and in other high resolution CW RRS
experiments~\cite{SimonetAl05CPL} for the outer walls of
double-walled carbon nanotubes, individual SWNTs on substrates,
and individual SWNTs dispersed in solution were approximately
5-6~cm$^{-1}$.  This observation is consistent with results of
previous studies on coherent phonons in
semiconductors~\cite{YeeetAl02PRL,LeeetAl03JAP} and clearly
indicates that the two techniques are sensitive to different
dynamical quantities.  CP data contains information on how
impulsively-generated transient coherent phonons decay.  Once
the coherent phonons are generated, their decay processes are
most likely to be through disappearance into lower-frequency
phonon modes by
anharmonicity~\cite{Merlin97SSC,Aku-LehetAl05PRB},
independent of any electronic states or processes.  Raman
linewidths are affected by the electronic states involved as
well as electron-phonon interactions.  This makes CP an ideal
method for phonon spectroscopy with precise determination of
both frequencies and decay times.

\begin{figure}
\includegraphics [scale=0.6] {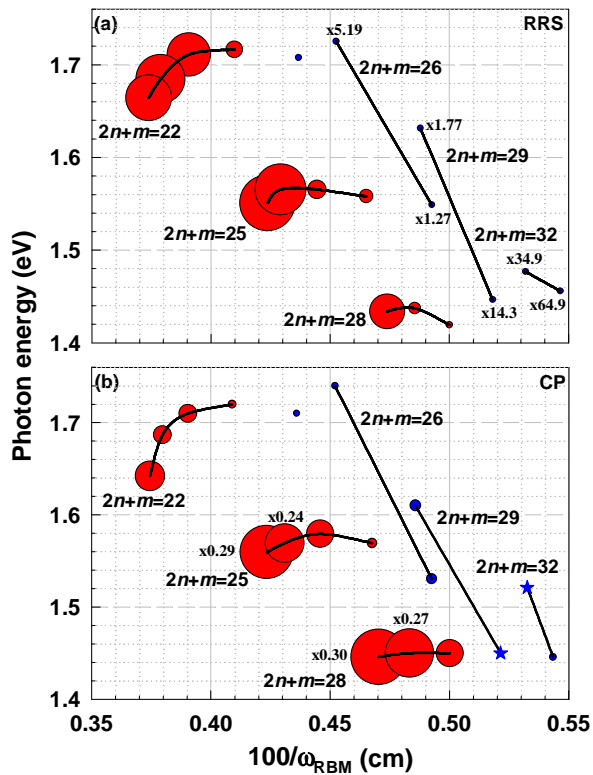}
\caption{
(a) Raman intensity and (b) coherent phonon signal as a function
of excitation energy and 100/$\omega_{\rm RBM}$, where
$\omega_{\rm RBM}$ is the phonon frequency.
Red (blue) circles denote nanotubes that satisfy
($n$$-$$m$) mod~3 = $-$1 (+1).  The black
lines connect members of the same (2$n$+$m$) family.  The
diameter of the circles is proportional to measured
intensities~\cite{derivative}.
The blue stars in (b) indicate chiralities with
large uncertainty in their determined excitonic energies and
coherent phonon signal. } \end{figure}

Figure 3 summarizes the observed various RBMs both for CW Raman
scattering [Fig.~3(a)] and coherent phonons
[Fig.~3(b)]~\cite{derivative}. The dependence of
CP signal strength as a function of photon energy and phonon
frequency exhibits several of the same trends both predicted
theoretically~\cite{PopovetAl04NL,MachonetAl05PRB,Goupalov1,JiangetAl05PRB,Goupalov2}
and measured
experimentally~\cite{DoornetAl04APA,MaultzschetAl05PRB2} by CW
Raman scattering intensity as a function of excitation energy
and Raman shift. First, when exciting the second excitonic
transition for an individual semiconducting SWNT,
$E_{22}$, ($n$$-$$m$) mod 3 = $-$1 nanotubes have markedly stronger
signal than ($n$$-$$m$) mod 3 = +1 nanotubes.  Second, within a
(2$n$+$m$) nanotube family, signal strength increases from
nanotubes with large chiral angles (near-`armchair') to
nanotubes with small or 0$^{\circ}$ chiral angle (`zigzag').
However, the two techniques diverge in similarity when comparing
the overall strength in signal.  Namely, in Raman scattering,
signal strength decreases as (2$n$+$m$) = constant increases
whereas CP signal increases with increasing (2$n$+$m$) =
constant.  We currently have no theoretical model to explain
this striking difference between CP measurements and RRS, and it
is thus open for further studies.

\begin{figure}
\includegraphics [scale=0.75] {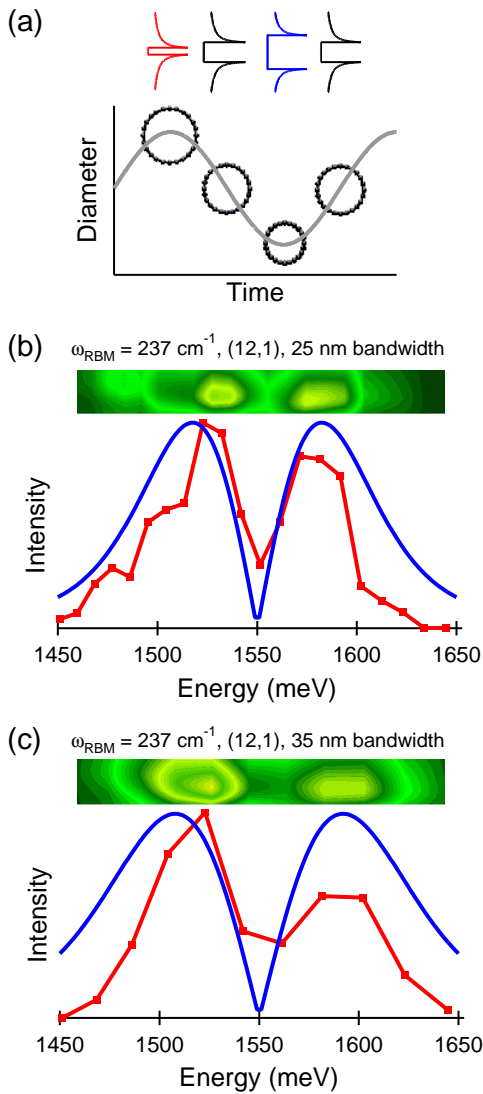}
\caption{(a)~Time-dependent band gap due to the radial-breathing
mode of coherent lattice oscillations.
(b) and (c): The photon
energy dependence of the coherent phonon signal intensity
(both contour and 2D plots) for the (12,1) tube
with a probe bandwidth of (b)~25~nm and (c)~35~nm together with
theoretical curves (blue sold lines).
} \end{figure}

Finally, we discuss how the generation of CPs of RBMs modifies
the electronic structure of SWNTs and how it can be detected as
temporal oscillations in the transmittance of the probe beam.
The RBM is an isotropic vibration of the nanotube lattice
in the radial direction, i.e., the diameter ($d_{\rm t}$)
periodically oscillates at frequency $\omega_{\rm RBM}$.  This
causes the band gap $E_{\rm g}$ to also oscillate at
$\omega_{\rm RBM}$ [see Fig.~4(a)] because $E_{\rm g}$ directly
depends on the nanotube diameter (roughly $E_{\rm g} \propto
1/d_{\rm t}$).  As a result, interband transition energies
oscillate in time, leading to ultrafast modulations of optical
constants at $\omega_{\rm RBM}$, which naturally explains the
oscillations in probe transmittance.  Furthermore, these
modulations imply that the absorption coefficient at a fixed
probe photon energy is modulated at $\omega_{\rm RBM}$.
Correspondingly, the photon energy dependence of the CP signal
shows a {\em derivative-like} behavior, \`a la
modulation spectroscopy~\cite{Cardona69Book} [see
Figs.~4(b) and 4(c)].  We modeled this behavior assuming that
the CP signal intensity is proportional to the absolute value of
the convoluted integral of the first derivative of a Lorentzian
absorption line and a Gaussian probe beam profile.  The results,
shown as solid lines in Figs.~4(b) and 4(c), successfully
reproduce the observed double peaks, whose energy separation
changes with the bandwidth of the probe.  The {\em symmetric}
double-peak feature confirms the excitonic nature of the
absorption line, in contrast to the asymmetric shape expected
from the 1-D van Hove singularity.

In addition to revealing a novel optical process, involving
both 1-D excitons and phonons simultaneously, this study opens up
a number of new possibilities to study SWNTs. In particular, it
has many advantages over conventional CW characterization
methods, including i)~easy tuning of the center wavelength of
the pump pulse, ii)~simultaneous excitation of multiple
vibrational modes due to the broad spectrum of the exciting
pulse, iii)~no Rayleigh scattering background at low
frequency, iv)~no photoluminescence signal, and v)~direct
measurement of vibrational dynamics including its phase
information and dephasing times.  The excitonic nature of
interband optical excitation manifests itself in the detection
process for the CP oscillations by the coupling with the broad
spectrum of probe pulses.  Furthermore, the expansion of the
spectrum of femtosecond pulses into deep mid-infrared and
visible ranges will allow us to explore large-diameter carbon
nanotubes and metallic carbon nanotubes.  Finally, the ability
of CP measurements to trace the first derivative of the
excitonic absorption peaks of specific chirality ($n$,$m$) will
allow in-depth study of the lineshape of these resonances.

YSL acknowledges financial support from Konkuk University.
This work was supported in part by the Robert A. Welch
Foundation (No.~C-1509) and NSF (Nos.~DMR-0134058 and
DMR-0325474). SKD appreciates the support of the LANL Integrated
Spectroscopy Laboratory and partial financial support of this
work from the LANL LDRD program.  We thank C.~Kittrell for
helpful discussions and comments.

\bigskip

\noindent$^{\dagger}$To whom correspondence should be addressed.
Electronic address: kono@rice.edu.


\begin{thebibliography}{25}
\bibitem{derivative}
The CP amplitude in 3(b) was calculated by taking the peak height
of the maxima of the probe profile, assuming it occurs at the
inflection point of a Lorentzian function centered at the second
excitonic transition for a given ($n$,$m$) semiconducting
nanotube with a half width at half maximum of 30~meV, and
setting it equal to the expression for the first derivative of
the Lorentzian function.
\end{thebibliography}

\end{document}